\documentclass[amsmath,groupedaddress,showpacs,amssymb,aps,pra,twocolumn]{revtex4}
\usepackage{graphicx}
\usepackage{amsmath}
\newcommand{\figref}[1]{Fig.~\ref{fig:#1}}
\newcommand{\equref}[1]{Eq.~(\ref{eq:#1})}
\newcommand{\secref}[1]{section~\ref{sec:#1}}

\newcommand{\quarter}{\frac{1}{4}}
\newcommand{\half}{\frac{1}{2}}

\newcommand{\ud}{\mathrm{d}}

\newcommand{\operator}[1]{\hat{\mathcal{#1}}}

\newcommand{\bra}[1]{\langle #1 |}
\newcommand{\ket}[1]{| #1 \rangle}
\newcommand{\bracket}[2]{\langle #1 | #2 \rangle}

\newcommand{\halfket}[3]{\ket{#1^{(#2)}_\half(#3)}}

\begin{document}
\title{Violation of local realism in a high-dimensional two-photon setup with
non-integer spiral phase plates}
\author{S.S.R. Oemrawsingh}
\author{A. Aiello}
\author{E.R. Eliel}
\author{G. Nienhuis}
\author{J.P. Woerdman}
\affiliation{Huygens Laboratory, Leiden University\\
P.O.\ Box 9504, 2300 RA Leiden, The Netherlands}
\begin{abstract}
We propose a novel setup to investigate the quantum non-locality of orbital
angular momentum states living in a high-dimensional Hilbert space.  We
incorporate non-integer spiral phase plates in spatial analyzers, enabling us
to use only two detectors. The resulting setup is somewhat reminiscent of that
used to measure polarization entanglement. However, the two-photon states that
are produced, are not confined to a $2\times 2$-dimensional Hilbert space, and
the setup allows the probing of correlations in a high-dimensional space. For
the special case of half-integer spiral phase plates, we predict a violation
of the Clauser-Horne-Shimony-Holt version of the Bell inequality ($S\leq2$),
that is even stronger than achievable for two qubits
($S=2\sqrt{2}$), namely $S=3\frac{1}{5}$.
\end{abstract}
\maketitle
\section{Introduction}
Recently, the orbital angular momentum (OAM) of light has drawn considerable
interest in the context of quantum information processing. The spatial degrees
of freedom involved in OAM \cite{WOE92} provide a high-dimensional
alphabet to quantum information processing (i.e. qu$N$its instead of
qubits) \cite{ZEI01,ZEI02:2}.  Additionally, since OAM is associated with the
topology of the electromagnetic field, the use of this observable in quantum
entanglement may lead to states that are inherently robust against
decoherence \cite{PRE99}.

The most popular OAM analyzer when dealing with conservation, correlation and
entanglement of OAM consists of a so-called fork hologram \cite{BRA99}, i.e. a
binary phase hologram containing a fork in its center \cite{VAS93}, together
with a spatial-mode detector consisting of a single-mode fiber connected to a
single-photon detector; such analyzers have been used in the three-dimensional
case, i.e.  $N=3$, by Vaziri~\emph{et~al.}, \cite{ZEI02:2}. In that experiment,
proof of entanglement of the OAM degree of freedom of two photons was given by
showing that a generalized Bell inequality was violated; this scheme requires
6 detectors, namely 3 in each arm, and one has to measure $3\times 3$
coincidence count rates \cite{ZEI02:2} to perform a measurement for a single
setting of the analyzers.

In the present paper, we consider the use of spiral phase plates
(SPPs) \cite{WOE94} instead of phase holograms in an OAM entanglement
setup, enabling us to investigate high-dimensional entanglement with only
\emph{two} detectors.  More specifically, we will consider SPPs that impose on
an optical beam a \emph{non-integer} OAM expectation value per photon, in
units of $\hbar$ \cite{WOE94}.  With such devices, combined with
single-mode fibers to form quantum-state analyzers, we propose to build an OAM
entanglement setup that is reminiscent of the usual setup to measure
polarization entanglement \cite{KWI95}, where the rotational settings of the
analyzers (polarizers in that case) is varied. We will show that it is
possible to identify SPP analyzer settings, in the spirit of horizontal and
vertical aligned polarizers, when using \emph{half}-integer SPPs, allowing
observation of high-dimensional entanglement ($N>2$), in contrast to the
polarization case ($N=2$).  These claims are supported by
calculations; we predict highly non-classical quantum correlations
($S=3\frac{1}{5}$), i.e.  stronger quantum correlation between two photons
than the maximum correlation between two qubits ($S=2\sqrt{2}$).

\section{Spiral phase plates}
\begin{figure}
\centerline{\includegraphics[width=8cm]{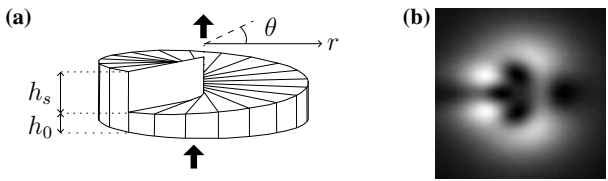}}
\caption{(a) Schematic drawing of a SPP. The device shifts the
phase of an incident beam proportional to the azimuthal angle $\theta$.
(b) A calculated far-field diffraction pattern of a fundamental Gaussian beam
after propagating through an $\ell=3\half$ plate positioned in its waist plane,
showing that rotational symmetry is broken. Black and white denote low and high
intensity, respectively.}\label{fig:spp}
\end{figure}

A SPP, shown in \figref{spp}(a), is a transparent dielectric plate with a
thickness that varies as a smooth ramp, thus phase shifting an incident field
linearly with the azimuthal angle $\theta$ \cite{WOE94}. As can be seen,
the plate carries a screw discontinuity, expressed by the spiraling
thickness, and an edge discontinuity, i.e. a radially oriented
step with height $h_s$.
The difference between the maximum and minimum phase shift is written as
$2\pi\ell$, where $\ell$ is not necessarily integer; in fact, $\ell$ depends
on depends on the step height $h_s$, the difference in refractive indices of
the SPP and surrounding medium, and the wavelength of the incident
light \cite{WOE94}. Thus, a photon propagating through this plate will acquire
an OAM with expectation value equal to $\ell\hbar$ \cite{NIE92,WOE94}.  Placing
such a plate in the waist of Laguerre-Gaussian beam, the field in the
exit plane just behind the plate will be described by
\begin{equation}\label{eq:sppoperator}
\bra{r,\theta}\operator{S}(\ell)\ket{l,p}=
u_\mathrm{LG}^{lp}(r,\theta)
\exp\left(i\ell\theta\right),
\end{equation}
where $\ket{l,p}$ are the Laguerre-Gaussian (LG) field states and
$\operator{S}(\ell)$ is the operator representing the effect of the SPP on the
input mode. Note that we have neglected the uniform phase shift that is caused
by the SPP's base with height $h_0$, since it acts as a plane-parallel plate.
By adding another plane-parallel plate with the appropriate thickness, the
total phase shift of these two can be made equal to an integer multiple of
$2\pi$.

The function $u_\mathrm{LG}^{lp}(r,\theta)$ in \equref{sppoperator} is the
complex amplitude of the Laguerre-Gaussian beam in its waist plane, given
by \cite{WOE92}
\begin{equation}\label{eq:lgfunction}
\begin{split}
\bracket{r,\theta}{l,p}=&u^\mathrm{LG}_{lp}(r,\theta)\\
=&
C_{lp}
\left(-1\right)^p
\left(\frac{r\sqrt{2}}{w_0}\right)^{|l|}
L^{\left|l\right|}_p\left(\frac{2r^2}{w_0^2}\right)\\
&\times\exp\left(-\frac{r^2}{w_0^2}\right)\exp\left(il\theta\right),
\end{split}
\end{equation}
where $w_0$ is the waist radius, $L^l_p(x)$ an associated Laguerre
polynomial \cite{ABR65} and $C_{lp}$ a normalization constant.  In the paraxial
limit, the LG free-space modes, enumerated by the (integer) indices $l$ and
$p$, form a complete basis of spatial modes. The LG index $l$, which should
not be confused with the SPP step index $\ell$, is related to the OAM that is
carried in that LG mode, namely $l\hbar$ per photon \cite{WOE92,NIE92,WOE94},
while the index $p$ provides information on the number of nodes along a
transverse radius of the mode.  When placing a SPP in a beam in a single LG
mode, the output will generally be in a \emph{superposition} of LG modes and
will be \emph{no} longer \emph{invariant} under free-space propagation.  For
integer values of the step index $\ell$ of the SPP, the edge discontinuity is
effectively absent and this superposition will only be with respect to the
LG-index $p$ \cite{WEG92,WOE94}.  In that case, the intensity distribution of
the mode will be doughnut-shaped in the far field.  For non-integer $\ell$
values, the edge discontinuity does not vanish and the
superposition of modes will also be with respect to the index $l$, which
is related to the OAM of the mode \cite{WOE94}. Such superpositions consist in
principle of an infinite number of LG components.  Effectively, this number is
finite and increases with $\ell$; as an example, if $\ell=\half$, 11 LG
components are sufficient to describe 87\% of the field behind the SPP, while
for $\ell=\frac{5}{2}$, 224 LG components are required.  As SPPs with
non-integer $\ell$ can \emph{create} such high-dimensional superpositions of
OAM modes, we anticipate that, when employed suitably, they can also
\emph{project} onto such superpositions.  We therefore propose to incorporate
such non-integer SPPs in analyzers for OAM states living in a high-dimensional
Hilbert space.

From a topological point of view, SPPs with non-integer $\ell$ imprint a mixed
screw-edge dislocation on an incident field. The result is rotational
asymmetry of the imprinted phase distribution and thus of the emerging field,
which becomes visible in the far-field intensity profile (\figref{spp}(b)).  It
is the orientation of the step in the transverse plane, that we wish to
exploit as an analyzer setting in a new bipartite entanglement scheme.

Since $\ell$ shall be chosen to have a non-integer value, it is important to
realize that, when an incident field passes through an SPP in combination with
its \emph{complement} (i.e. a SPP with the same step height and orientation,
but an \emph{inverted} vorticity), the beam basically passes through a
plane-parallel plate that shifts the phase of the field, in an azimuthally
uniform way, by $2\pi\ell$,
\begin{equation}
\operator{S}^\mathrm{compl}(\ell)\operator{S}(\ell)=
\exp\left(i2\pi\ell\right)\hat I,
\end{equation}
where $\hat I$ is the identity operator and where we keep the exponent since
$\exp\left(i2\pi\ell\right)\ne 1$ for non-integer $\ell$. Since
$\operator{S}(\ell)$ is unitary, it follows that
\begin{equation}
\operator{S}^\mathrm{compl}(\ell)=\exp\left(i2\pi\ell\right)
\operator{S}^\dag(\ell),
\end{equation}
where $\operator{S}^\dag(\ell)$ is the Hermitean conjugate of
$\operator{S}(\ell)$.  As $\operator{S}^\dag(\ell)$ and
$\operator{S}^\mathrm{compl}(\ell)$ only differ by a multiplicative phase
factor we can again use a plane-parallel plate in the experiment to compensate
for this phase factor. Similar to the uniform phase shift caused by the SPP's
base $h_0$, we will neglect this phase shift as well, as it can be trivially
dealt with. Henceforth, the operator for the compensating SPP, with inverted
vorticity, will be represented by the operator $\operator{S}^\dag(\ell)$.

\section{Proposal for an experiment}
\begin{figure}
\centerline{\includegraphics[width=8cm]{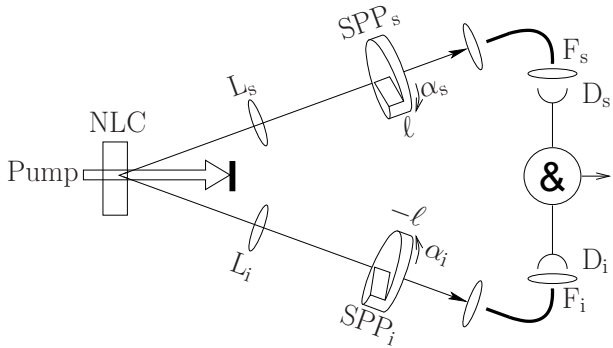}}
\caption{Proposed experimental setup.  A nonlinear crystal (NLC) splits a pump
photon in a signal photon and an idler photon by the process of SPDC.
In each path, a SPP (SPP$_\mathrm{s,i}$) is
inserted with a single-mode fiber F$_\mathrm{s,i}$, together forming the
analyzer.  The coincidence count rate of detectors D$_\mathrm{s,i}$ is
measured as a function of the SPP angular settings $\alpha_\mathrm{s}$ and
$\alpha_\mathrm{i}$.}\label{fig:setup}
\end{figure}

In the experiment on OAM qutrits ($N=3$), fork holograms were used
\cite{ZEI02:2}. Those holograms can only modify the OAM expectation value by
an \emph{integer} number \cite{SOS98footnote}, depending on the diffraction
order of the hologram. It also required the use of \emph{three} analyzers in
each arm of a spontaneous parametric down-conversion (SPDC) setup. In contrast, our proposed OAM entanglement
experiment uses \emph{non-integer} SPP state detectors and only requires a
\emph{single} analyzer-detector combination in each arm.  This scheme is shown
in \figref{setup}; it has been inspired by the setup used in polarization
entanglement \cite{KWI95}. The SPPs that are inserted in the two arms should
be chosen to obey conservation of OAM (see also \secref{entanglement}). Thus,
if the classical pump beam does not contain any OAM, a typical choice for the
signal and idler SPPs would be to have step indices $\ell$ and $-\ell$
respectively, where, in our case, $\ell$ should have a non-integer value. The
role of the lenses in signal and idler path, L$_\mathrm{s}$ and L$_\mathrm{i}$
respectively, will be discussed in \secref{entanglement}. Finally, single-mode
fibers are indicated by F$_\mathrm{s}$ and F$_\mathrm{i}$.

By manipulating the transverse axes of our analyzers, namely the SPP steps, we
have access to various photon states that live in a high-dimensional OAM
Hilbert space, as mentioned earlier. We may orient these edges arbitrarily in
the transverse plane, as shown in \figref{setup}, thus allowing their use as
angular analyzers. When combined with a single-mode fiber, the SPP with step
index $\ell$, set at an azimuthal angle $\alpha_\mathrm{s}$, projects the
incident photon state out onto the OAM state with expectation value $-\ell$
with edge angle $\alpha_\mathrm{s}$. As we will argue in
\secref{entanglement}, the coincidence count rate will depend, like in
polarization entanglement, only on the \emph{relative} angle of the transverse
axes.

We stress that, in spite of the superficial similarity between a polarizer and
a non-integer SPP, they are of course very different devices; for example,
while polarization corresponds to \emph{alignment}, the SPP edge corresponds
to \emph{orientation}. In other words, with a polarizer, one can analyze the
alignment of the electrical field oscillation, either horizontal or vertical,
thus yielding a periodicity of $\pi$ when rotated. With a SPP with non-integer
$\ell$ one can analyze the spatial orientation of a field, thus yielding a
periodicity of $2\pi$ when rotated (see e.g. \figref{spp}(b)).  Thus we expect
that the coincidence count rate will have a periodicity of $2\pi$ when one of
our analyzers is rotated.

An equally important difference between the two cases is that, whereas
polarization Hilbert space is two dimensional, OAM Hilbert space is infinite
dimensional. As we will see, this makes rotation of SPPs fundamentally
different from rotation of a polarizer. In order to address these aspects
explicitly, we need  a basis of the OAM Hilbert space that is suited for our
purpose.

\section{Non-integer OAM states}
Our aim is to construct a complete, orthonormal basis that contains
non-integer OAM states as basis elements. To this end, we consider all states
in the polar representation and separate the radial and angular parts, so that
the arbitrary state $\ket{\vec{r}}=\ket{r,\theta}$ can be written as $\ket{r}\ket{\hat{r}}$,
where we have introduced the \emph{direction} (i.e. angular) ket
$\ket{\hat{r}}$. This allows us to introduce a complete basis set of angular
states, the eigenstates $\ket{l}$ of the OAM operator $\operator{L}_z$, so
that $\operator{L}_z\ket{l}=l\ket{l}$. A Laguerre-Gaussian state can thus be
separated into a radial and angular part, so that
$\ket{l,p}=\ket{\rho_{lp}}\ket{l}$, where $\ket{\rho_{lp}}$ is the radial part of the
state, which is of less importance in this paper. The OAM eigenstates can be
transformed from integer to non-integer OAM states.
By applying a unitary operator to the integer OAM basis, its
completeness and orthonormality are conserved; the unitary operator that we
use is the SPP operator, introduced in \equref{sppoperator}:
\begin{equation}\label{eq:nonintfuncs}
\bra{\hat r}\operator{S}(\lambda)\ket{l}\equiv
\bracket{\hat
r}{a^{(l)}_\lambda}=\frac{1}{\sqrt{2\pi}}\exp{i(l+\lambda)\theta}.
\end{equation}
The new basis $\left\{\ket{a^{(l)}_\lambda}\right\}$ has its components
enumerated by $l$, each with OAM equal to $(l+\lambda)\hbar$, where
$\lambda\in[0,1)$  is a constant (not to be confused with the wavelength of
the light).  Different values of $\lambda$ define different bases of OAM
Hilbert space, each basis being complete and orthonormal.

As an example, passing a photon in a fundamental Gaussian mode
$\ket{0,0}=\ket{\rho_{00}}\ket{0}$ through a SPP with $\ell=\frac{2}{3}$, creates
the state $\ket{\rho_{00}}\ket{a^{(0)}_{2/3}}$; sending this latter photon
through a SPP with an aligned step and $\ell=-\frac{2}{3}$, combined with a
single-mode fiber, then yields a detection probability of unity.

\section{Orientation of the edge dislocation}\label{sec:orientation}
\begin{figure}
\centerline{\includegraphics[width=8cm]{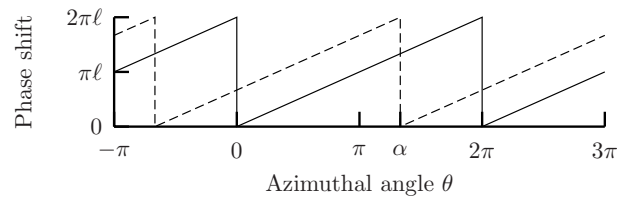}}
\caption{Plot of the phase shift imprinted by a SPP as a
function of the azimuthal angle. The solid line shows this shift for a
plate with its edge oriented at $\theta=0$, while the dashed line shows the
phase shift for a plate with its edge at $\theta=\alpha$.}\label{fig:rotation}
\end{figure}

Since we intend to rotate the non-integer SPPs, it is important to also have a
description of the states that arise when the SPP is rotated while the
coordinate system is fixed.  To gain insight, we show in \figref{rotation} the
phase shift imposed by a SPP as a function of the azimuthal angle.
The solid line represents the imprinted phase for a SPP with its edge oriented
at an angle $\theta=0$, while the dashed line corresponds to a SPP with edge
orientation $\theta=\alpha$. We can now generalize the definition of the
operator $\operator{S}(\ell)$ to the operator $\operator{S}(\alpha,\ell)$,
which includes the orientation $\alpha$ of the edge dislocation. We find
\begin{equation}\label{eq:orientop}
\begin{split}
\bra{\hat r}\operator{S}(\alpha,\ell)\ket{l}
=&\frac{1}{\sqrt{2\pi}}\exp\left[{i(l+\ell)\theta}\right]\\
&\times\left\{
\begin{array}{l @{\quad} l}
\exp\left[i(2\pi-\alpha)\ell\right], & 0\leq\theta < \alpha,\\
\exp\left(-i\alpha\ell\right), & \alpha\leq\theta < 2\pi,
\end{array}
\right.
\end{split}
\end{equation}
where $\alpha,\theta\in[0,2\pi)$. This also allows us to generalize the states
$\ket{a^{(l)}_\lambda}$ with orientation $\theta=0$ to states with arbitrary
orientation, $\ket{a^{(l)}_\lambda(\alpha)}$. From \equref{orientop}, it is
immediately clear that, when neglecting the uniform phase shift, the
complementary SPP operator $\operator{S}^\mathrm{compl}(\alpha,\ell)$ is equal
to $\operator{S}(\alpha,-\ell)=\operator{S}^\dag(\alpha,\ell)$.

\begin{figure}
\centerline{\includegraphics[width=8cm]{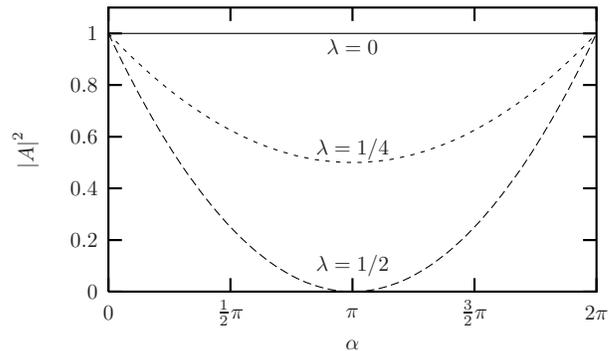}}
\caption{The overlap (see \equref{weight}) between a non-integer OAM state
and the identical state rotated over $\alpha$.  When $\lambda=0$, the states
are identical aside from a trivial phase shift, due to the vanishing edge
dislocation. For half-integer OAM, i.e.  $\lambda=\half$, the two
states are generally different, leading to a parabolic dependence of their
overlap; the two states are orthogonal when
$\alpha=\pi$.}\label{fig:projection}
\end{figure}

Since the basis $\left\{\ket{a^{(l)}_\lambda(0)}\right\}$ is complete, the
states after rotation, $\ket{a^{(l)}_\lambda(\alpha)}$ can be written as a
superposition of these basis states. Thus the decomposition of
$\ket{a^{(l)}_\lambda(\alpha)}$ into the basis
$\left\{\ket{a^{(l)}_\lambda(0)}\right\}$ depends on the angle $\alpha$.  To
illustrate this, we make a projection of a non-integer state oriented at
$\theta=\alpha$, onto the same state with orientation $\theta=0$. For this, we
choose a SPP with step $\ell=j+\lambda$, where $j$ is the integer part of the
step (not to be confused with the integer LG index $l$), and
$\lambda\in[0,1)$, yielding the overlap amplitude
\begin{equation}\label{eq:projection}
\begin{split}
A_\lambda(\alpha)=&\bracket{a^{(l+j)}_\lambda(0)}{a^{(l+j)}_\lambda(\alpha)}\\
=&\bra{l}\operator{S}^\dag(0,j+\lambda)\operator{S}(\alpha,j+\lambda)\ket{l}\\
=&\frac{1}{2\pi}\left[2\pi-\alpha+\alpha\exp\left(i2\pi\lambda\right)\right]\\
&\times\exp\big[-i(l+j+\lambda)\alpha\big],
\end{split}
\end{equation}
where $\ket{l}$ is the OAM operator eigenstate with eigenvalue $l$. The
overlap probability is then
\begin{equation}\label{eq:weight}
|A_\lambda(\alpha)|^2=\left(1-\frac{\alpha}{\pi}\right)^2\sin^2\left(\lambda\pi\right)
+\cos^2\left(\lambda\pi\right),
\end{equation}
which depends neither on the integer part of the step index $j$ nor on the OAM
state $l$. For non-zero values of $\lambda$, this overlap function has a
quadratic dependence on the orientation $\alpha$.  This function is plotted in
\figref{projection} for various values of $\lambda$.  It illustrates that,
when $\lambda=0$, the projection does not change, as expected. For values of
$\lambda\ne 0$, the outcome of the projection is less trivial, with
$\lambda=\half$ providing an especially interesting case: when the state is
rotated over $\alpha=\pi$ by rotating the SPP, the state is orthogonal to the
non-rotated state.
We will call the half-integer states $\halfket{a}{l}{0}$ and
$\halfket{a}{l}{\pi}$ `up' and `down', respectively, referring to the
orientation of the edge part of the dislocation. In principle, this
orientational label can be used for \emph{any} non-integer OAM state, but for
half-integer OAM states it carries an analogy to fermionic spin, as our `up'
and `down' states are orthogonal just like up and down spin $\half$. In
polarization entanglement the comparable orientational labels are known as `H'
for horizontal and `V' for vertical polarization.  To make our proposed OAM
entanglement setup maximally equivalent to the polarizer setup, we focus from
now on mainly on the case $\lambda=\half$. Note that, as \equref{weight} is
independent of $j$, any SPP with $\ell=j+\lambda$ may be used, as long as
$\lambda=\half$.

There is, however, a key difference between the half-integer OAM states on the
one hand and  the fermionic spin states or polarization states on the other
hand. Angularly intermediate states of the fermionic spin or of polarization
are a superposition of the two orthogonal basis states `up' and `down', or `H'
and `V', so that the overlap function varies as $\sin^2\alpha$. However, from
\equref{weight} we see that in the present case, the dependence of $|A|^2$ on
$\alpha$ yields a \emph{parabola}. Thus we conclude that, as the SPP is
rotated from `up' to `down', the OAM state follows a path through Hilbert
space that is \emph{not} confined to the two-dimensional subspace spanned by
the `up' and `down' states.  Note that the OAM expectation value will be
conserved along this path, as rotating the SPP has, of course, no effect on
its step height, i.e. on $\ell$.

\section{Entanglement of half-integer orbital angular momentum
states}\label{sec:entanglement}
When dealing with OAM entanglement, issues of conservation of OAM during the
spontaneous parametric down-conversion process arise. This conservation
is discussed in many papers \cite{BAR00,WOE01,BAR01,ZEI01,ALL02,ARN02,TER03}
and it seems that OAM is indeed conserved if two conditions are fulfilled,
namely (i) the paraxial limit, and (ii) the thin-crystal limit. In practice
this is usually the case and we will therefore work within these limits.

The two-photon state can be described in the spatial polar representation as
\cite{NIE04}
\begin{equation}\label{eq:standardtwophoton}
\ket{\Psi}=\int\,\ud\vec{r}\,\mathcal{P}(\vec{r})\,\ket{\vec{r}}_1\,
\ket{\vec{r}}_2,
\end{equation}
where $\mathcal{P}(\vec{r})$ is the mode function of the pump beam.
We restrict ourselves to a pump beam in a pure OAM mode so that
$\mathcal{P}(\vec{r})$ can be written as $\mathcal{P}(r)\exp\left(iq\theta\right)/\sqrt{2\pi}$,
where $\mathcal{P}(r)$ is the radial part of the pump mode function. We can now
write the two-photon state in the half-integer SPP basis
$\left\{\halfket{a}{n}{0}\right\}$, thus yielding
\begin{equation}\label{eq:twophotonint}
\begin{split}
\ket{\Psi}=&\frac{1}{\sqrt{2\pi}}\int\,\ud r\,r
\mathcal{P}(r)\,\ket{r}_1\,\ket{r}_2\\
&\times\sum_{m,n=-\infty}^\infty\left[
\int\,\ud\theta\,\exp{\left(iq\theta\right)}
\bracket{a^{(n)}_\half(0)}{\hat{r}}\bracket{a^{(m)}_\half(0)}{\hat{r}}\right]\\
&\phantom{\times\sum_{m,n=-\infty}^\infty}\times\halfket{a}{n}{0}\halfket{a}{m}{0}\\
=&\frac{1}{\sqrt{2\pi}}\ket{R}\sum_n\halfket{a}{n}{0}\,\halfket{a}{q-n-1}{0},
\end{split}
\end{equation}
where we have defined the radial ket,
\begin{equation}
\ket{R}\equiv\int\,\ud r\,r\mathcal{P}(r)\,\ket{r}_1\,\ket{r}_2.
\end{equation}

We now continue to calculate the coincidence fringe that is expected in the
proposed experiment. For the pump beam, we shall assume a fundamental Gaussian
beam, so that $q=0$ in \equref{twophotonint}. In the signal path we place an
analyzer consisting of a SPP with $\ell=j+\half$ where $j$ is an integer, with
its orientation set to `up' ($\theta=0$), represented by
$\operator{S}(0,j+\half)$, and a single-mode fiber. When the detector clicks,
the signal state \emph{before} passing through this analyzer is collapsed to
$\operator{S}^\dag(0,j+\half)\ket{0,0}= \ket{\rho_{00}}\halfket{a}{-j-1}{0}$,
where $\ket{\rho_{00}}$ is the radial part of the fundamental Gaussian state.

Consequently, according to \equref{twophotonint} with $l=0$, the idler state
$\ket{\psi}_2$ is then collapsed to
\begin{equation}
\ket{\psi}_2=C\halfket{a}{j}{0},
\end{equation}
where
\begin{equation}
C=\frac{1}{\sqrt{2\pi}}\left(\bra{\rho_{00}}_1\bra{\rho_{00}}_2\right)\ket{R}.
\end{equation}
Thus when analysing the idler state with
$\operator{S}(\alpha,j+\half)\ket{0,0}$ we obtain the projection
\begin{equation}\label{eq:coincidence}
B(\alpha)=C \bracket{a^{(j)}_\half(\alpha)}{a^{(j)}_\half(0)}.
\end{equation}
Note that \equref{coincidence} has, aside from a prefactor, exactly the same
appearance as \equref{projection} with $\lambda=\half$. The coincidence fringe
is then given by the modulus squared,
\begin{equation}
\left|B(\alpha)\right|^2=\left|C\right|^2 \left(1-\frac{\alpha}{\pi}\right)^2,
\end{equation}
which is proportional to
\equref{weight}: we find a parabolic coincidence fringe.

The above reasoning to obtain the coincidence fringe
$\left|B(\alpha)\right|^2$ is valid for any choice of the signal SPP
orientation and only depends on the \emph{relative} orientation $\alpha$ of
the signal and idler SPPs. Thus a coincidence measurement on entangled OAM
pairs using half-integer OAM analyzers, will bring forth a coincidence fringe
that is parabolic, regardless of the \emph{individual} settings of the
analyzers.

\section{The CHSH version of the Bell inequality}
There have been several theoretical papers that address the generalization of
the Bell inequality \cite{BEL64} to quantify the violation of local realism of
two $N$-dimensional particles
(qu$N$its) \cite{MER80,PER92,ZEI00:2,ZUK01,ZEI02:2,POP02}; an example of a
qu$N$it is a spin-$s$ particle with $2s+1=N$. It has been pointed out that in
this case the use of $m_s$-sorting devices, such as Stern-Gerlach analyzers,
does \emph{not} offer access to higher-dimensional quantum correlations,
presumably because the action of a Stern-Gerlach analyzer depends only on the
alignment of its quantization axis \cite{ZEI00:2}. Instead of larger spin
values ($s>\half$), the use of spatial degrees of freedom together with Bell
multiports has been advocated to gain access to the multidimensional aspects
of entanglement \cite{ZEI00:2}. We stress that all this is different from our
proposed use of half-integer SPPs ($s+\half$).  These devices do \emph{not}
produce finite-$N$ qu$N$its, but imprint the infinite OAM dimensionality of
the (oriented) edge on a transmitted light field.  Rotation of this edge is
equivalent to a partial exploration of the complete Hilbert space along a
certain path, namely an iso-OAM path; due to this complexity, it is not clear
how a generalized Bell inequality could be applied to our case.

However, instead of using a generalized high-dimensional bipartite Bell
inequality, it is allowed to use an inequality for lower-dimensional
two-particle entanglement \cite{POP02}. Thus we choose, as in the polarization
case, the inequality introduced by Clauser, Horne, Shimony and Holt (CHSH) for
a measurement where the coincidence probability is expected to be a function
of only $\alpha_\mathrm{s}-\alpha_\mathrm{i}$ \cite{HOL69}. When relabelling
$\alpha_\mathrm{s}$ and $\alpha_\mathrm{i}$ as $\alpha_1$ and $\alpha_2$ (in
no particular order), the CHSH inequality is given by \cite{PER93,KWI95}
\begin{equation}\label{eq:sparameter}
\begin{split}
S=&E(\alpha_1,\alpha_2)-
  E(\alpha_1',\alpha_2)+E(\alpha_1,\alpha_2')\\
 &+ E(\alpha_1',\alpha_2')\leq 2.
\end{split}
\end{equation}
The function $E$ is specified, for the variables $x,y$,
as \cite{RAP81,KWI95}
\begin{equation}\label{eq:eparameter}
E(x,y)=
\frac{P(x,y)+P(x^\perp,y^\perp)-P(x,y^\perp)-P(x^\perp,y)}{P(x,y)+P(x^\perp,y^\perp)+P(x,y^\perp)+P(x^\perp,y)}.
\end{equation}
The notation $x^\perp$ (and similarly for $y^\perp$) is used to indicate an
analyzer setting that analyses a state orthogonal to the state with setting
$x$. Thus in our case, $x^\perp\equiv x+\pi$ and $y^\perp\equiv y+\pi$.
$P(x,y)$ is the coincidence probability function, which is equal to
\begin{equation}
P(x,y)=\left|B\left(\left|y-x\right|\right)\right|^2=C^2\left(1-\frac{\left|y-x\right|}{\pi}\right)^2.
\end{equation}
As the periodicity in the present case is half that of the case of polarization
entanglement, we use the standard analyzer settings for polarization
entanglement \cite{HOR74,KWI95} multiplied by a factor of two:
$\alpha_1=-\quarter\pi$, $\alpha_1'=\quarter\pi$, $\alpha_2=-\half\pi$,
$\alpha_2'=0$.

Substitution yields a Bell parameter $S=3\frac{1}{5}$. This is the key result
of our paper; it indicates that in the case of entanglement of half-integer
OAM states, the maximum violation of the CHSH inequality, given by
\equref{sparameter}, is \emph{stronger} than the maximum violation that is
allowed in polarization entanglement, namely $S=2\sqrt{2}$.  In other words,
quantum non-locality of the photons in the proposed setup is stronger than the
maximum achievable for two qubits. To achieve this, only \emph{two} detectors
are required and only \emph{one} coincidence count rate is measured per analyzer
setting, in contrast to the OAM qu$N$it setup requiring $N$ detectors and
$N^2$ coincidence count rates per analyzer setting \cite{ZEI00:2,ZEI02:2}.

\section{Conclusions}
In this paper we have put forward a novel approach to demonstrate
high-dimensional entanglement of orbital angular momentum states. The proposed
setup uses analyzers that consist of non-integer SPPs and single-mode fibers,
enabling detection of high-dimensional entanglement with only two detectors.

The key idea is to use the orientation of the edge dislocation in the SPPs.
We specialize to the case of half-integer $\ell$, so that the orientation
of the edge as an analyzer setting can, to a certain extent, be treated
similarly as the axis of a polarizer in polarization entanglement. Instead of
horizontal and vertical polarization states, we deal with `up' and `down'
states, referring to the orientation of the edge dislocation. We analytically
calculate the coincidence fringe in the
entanglement setup and find it to be parabolic in shape, and periodic over
$2\pi$. When evaluating the well-known CHSH Bell parameter, we find
$S=3\frac{1}{5}$, i.e. we predict beyond-Bell pairing of two photons.  To
achieve this, we require only \emph{two} detectors, as opposed to the standard
multiport approach \cite{ZEI00:2,ZEI02:2}. This economic exploitation of the
spatial degrees of freedom seems to be a consequence of the singular nature of
our half-integer SPPs, which implies, in principle, infinite dimensionality.

Experimental verification of the outlined proposal is under way.

\section*{Acknowledgements}
We acknowledge M.~P.~van~Exter for fruitful discussions regarding the CHSH
version of the Bell inequality. This work is part of the research program of
the `Stichting voor Fundamenteel Onderzoek der Materie (FOM)' and is supported
by the EU programme ATESIT.

\section*{APPENDIX: PURE EDGE DISLOCATION}
In the present paper, we have seen that, for spiral phase plates with
half-integer step index $\ell$, i.e. imprinting a mixed screw-edge
dislocation, the coincidence fringe in a twin-photon experiment is parabolic
as a function of the relative orientation of the two radial edges, with
visibility equal to $1$. This then results in a value of the CHSH-Bell
parameter equal to $S=3\frac{1}{5}$.

An intriguing question is whether there are other devices with optical
singularities, besides the
half-integer spiral phase plate, that, in a quantum experiment, will give
rise to a similarly large value of the $S$-parameter. More precisely, can one
design a twin-photon experiment that results in a value of the $S$-parameter
larger than $2\sqrt{2}$, using devices that are simpler to produce and handle
than the spiral phase plate discussed so far?

\begin{figure}[!h]
\centerline{\includegraphics[width=6.2cm]{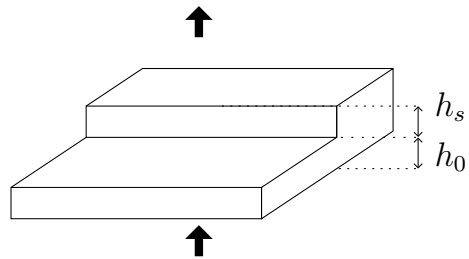}}
\caption{Similar to the spiral phase plate in \figref{spp}a), the
non-integer step phase plate has its step height $h_s$ chosen so that the phase
shift due to the thick part is a non-integer multiple of $2\pi$ with respect
to the thin part. The edge dislocation is chosen so that it goes through the
centre of the field mode.}\label{fig:qtheory:halfedgeplate}
\end{figure}

As we will see below, this question can be answered positively, and the
proposed device is surprisingly simple. It is a non-integer step phase plate
carrying a straight edge dislocation, as shown in
\figref{qtheory:halfedgeplate}.  Obviously, such a plate can be manufactured
more easily than the spiral phase plates, whose production and
characterisation is extensively discussed elsewhere \cite{WOE04,WOE04:2}.

Similar to the non-integer spiral phase plate, the non-integer step phase plate
contains an orientational degree of freedom, $\alpha$. It is possible to write
a unitary operator for the plate's action,
\begin{equation}\label{eq:qtheory:stepplateeq}
\bra{x,y}\operator{F}(\alpha,\phi)\ket{l} =
\frac{\exp\left(il\theta\right)}{\sqrt{2\pi}}
\times\left\{
\begin{array}{l @{\quad} l}
\exp\left(i\phi\right), & \alpha\leq\theta < \alpha+\pi,\\
1, & \\
\end{array}
\right.
\end{equation}
where $\alpha\in\left[0,\pi\right)$ represents the orientation of the step,
$\phi$ the optical phase delay resulting from the step,
$\theta\in\left[0,2\pi\right)$ the azimuthal angle and $\ket{l}$ the orbital
angular momentum eigenstates. For the special case $\phi=\pi$,
\equref{qtheory:stepplateeq} corresponds to the Hilbert transform
\cite{OPP75,KHO92}. The
non-integer step phase plate obviously also transforms the pure
orbital angular momentum states into high-dimensional states, due to the edge
dislocation. The definition as given here can be extended for
$\alpha\in\left[-\pi,\pi\right)$, seemingly complicating matters, but without
any impact on the calculations ahead.

Since the operator $\operator{F}$ is unitary, we can define a new,
complete basis $\left\{\ket{b^{(l)}_\phi(\alpha)}\right\}$, where
$\operator{F}(\alpha,\phi)\ket{l} = \ket{b^{(l)}_\phi(\alpha)}$. We can write
the states after rotation $\ket{b^{(l)}_\phi(\alpha)}$ in terms of
superpositions of the non-rotated states, $\ket{b^{(l)}_\phi(0)}$, where
the decomposition depends on the angle $\alpha$. We can illustrate this by
making a projection of such a state with $\alpha=0$ onto the same state
$\alpha \ne 0$. The overlap amplitude thus becomes
\begin{equation}
\begin{split}
A(\alpha)=&\bracket{b^{(l)}_\phi(0)}{b^{(l)}_\phi(\alpha)}\\
=&\frac{\alpha}{\pi}\left(\cos\phi-1\right) + 1,
\end{split}
\end{equation}
which is valid for values of $\alpha\in\left[-\pi,\pi\right)$. The overlap
probability is given by
\begin{equation}\label{eq:qtheory:hplateweight}
\left|A(\alpha)\right|^2=\left(\frac{\alpha}{\pi}\right)^2
\left(\cos\phi-1\right)^2 +2\frac{\alpha}{\pi}\left(\cos\phi-1\right)+1.
\end{equation}

\begin{figure}
\centerline{\includegraphics[width=8cm]{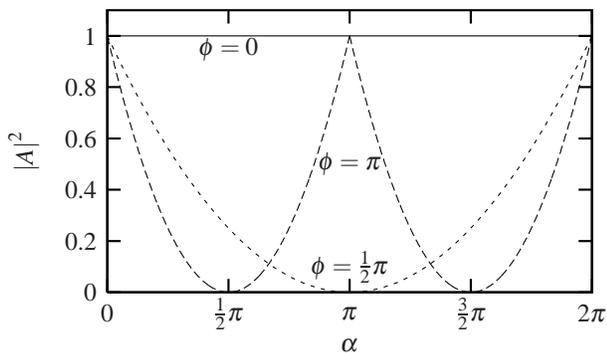}}
\caption{The overlap (see \equref{qtheory:hplateweight}) between a state with an
imprinted line dislocation, and the identical state rotated over $\alpha$.
When there is no dislocation ($\phi=0$), the states
are identical. For other phase differences $\phi$, the two
states are generally different; $\phi=\frac{1}{2}\pi$ and $\phi=\pi$ result
in a continuous parabolic fringe.}\label{fig:qtheory:hplateprojection}
\end{figure}

Figure~\ref{fig:qtheory:hplateprojection} shows the intensity fringe
as a function of $\alpha$ for three different values of $\phi$. For clarity,
we show the functions over the range $\alpha\in\left[0,2\pi\right)$ instead of
$\alpha\in\left[-\pi,\pi\right)$, which is allowed when $\alpha$ is periodic
over $2\pi$.
The most interesting cases occur when $\phi=\pi$ and $\phi=\pi/2$, when the
fringe becomes a parabola, periodic over $\pi$ and $2\pi$, respectively. The
result is then, aside from the periodicity, identical to that for a
half-integer spiral phase plate.

A plate with a single edge dislocation with phase difference $\pi$ or $\pi/2$
can thus be used, similar to a spiral phase plate, to investigate
entanglement. The setup would be identical to \figref{setup} with the
spiral phase plates replaced by the edge dislocation devices discussed here.
The calculation from two-photon state to coincidence fringe is the same as
discussed in \secref{entanglement}, using the states
$\ket{b^{(l)}_{\pi}(\alpha)}$, or
$\ket{b^{(l)}_{\pi/2}(\alpha)}$
instead of $\ket{a^{(l)}_{1/2}(\alpha)}$.

The CHSH-Bell inequality is violated maximally for these two-photon states,
using the set of sixteen angles as used in polarisation
entanglement for $\phi=\pi$, and the set as used for the
half-integer spiral phase plates for $\phi=\pi/2$.  Both cases yield an
CHSH-Bell parameter equal to $S=3\frac{1}{5}$.

\begin{figure}[!h]
\centerline{\includegraphics[width=8cm]{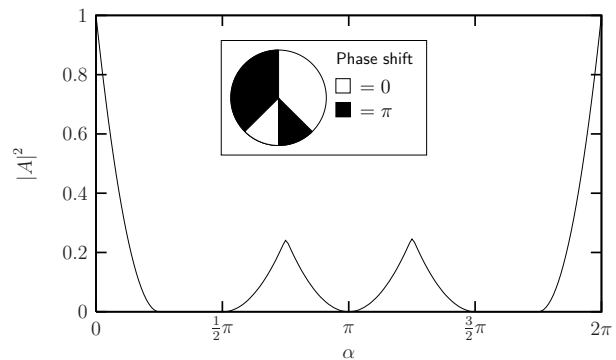}}
\caption{When using a plate as indicated in the inset, the overlap
(see \equref{qtheory:hplateweight}) as a function of $\alpha$ is as shown in
the graph. This results in $S=4$.}\label{fig:qtheory:peace}
\end{figure}

Finally, we note that binary phase plates only slightly more complex than the
type discussed above and pictured in \figref{qtheory:halfedgeplate}, can yield an even stronger
violation of the CHSH-Bell inequality. When using a plate with the shape as
indicated in the inset of \figref{qtheory:peace}, the maximum CHSH-Bell
parameter will be equal
to $S=4$, the highest value at all possible. For such a plate, we predict a
coincidence fringe with a shape as indicated in
\figref{qtheory:peace}.


\end{document}